\documentclass{iopjournal}
\usepackage{graphicx}
\usepackage{amsmath}
\usepackage{tabularx}

\begin{document}

\articletype{Topical Review} %	 e.g. Paper, Letter, Topical Review...

\title{Designing Quantum Matter in Pyrochlore Iridates: A Perspective on Recent Thin-Film Advances}

\author{Xiaoran Liu$^{1,*}$, Michael Terilli$^2$, Ana-Marija Nedić$^{3,4}$, Jiandong Guo$^{1,5}$, Jak Chakhalian$^2$}

\affil{$^1$Beijing National Laboratory for Condensed Matter Physics and Institute of Physics, Chinese Academy of Sciences, Beijing 100190, China}

\affil{$^2$Department of Physics and Astronomy, Rutgers University, Piscataway, NJ 08854, USA}

\affil{$^3$School of Physics and Astronomy, University of Minnesota, Minneapolis, MN, USA}

\affil{$^4$Department of Chemical Engineering and Materials Science, University of Minnesota, Minneapolis, MN, USA}

\affil{$^5$School of Physical Sciences, University of Chinese Academy of Sciences, Beijing 100049, China}

%\affil{$^*$Author to whom any correspondence should be addressed.}

\email{xiaoran.liu@iphy.ac.cn}


\begin{abstract}
%The pyrochlore iridates R$_2$Ir$_2$O$_7$ have emerged as a canonical platform for exploring exotic quantum phenomena arising from the intricate interplay of strong spin-orbit coupling, electron correlation, and geometric frustration. While bulk crystals of these materials have revealed a rich landscape of correlated and topological phases, recent breakthroughs in epitaxial thin-film synthesis and heterostructure engineering are unlocking an entirely new dimension of discovery. This perspective reviews these recent advancements, highlighting how new tuning knobs—such as dimensional confinement, epitaxial strain, and interfacial coupling—are being used to manipulate the delicate balance of competing interactions. We discuss several key discoveries enabled by this approach, including the realization of the magnetic Weyl semimetal phase in (111) films, the strain-engineering of magnetic multipolar orders, the emergence of a chiral spin-liquid-like state in the quasi-2D limit, and the discovery of novel electronic anisotropies at interfaces between pyrochlore iridates and other quantum materials like spin ice. These findings demonstrate that low-dimensional pyrochlore iridates provide a versatile and powerful platform for unraveling, controlling, and ultimately designing novel quantum states of matter. We conclude by outlining key open questions and future directions, from the synthesis of new heterostructures to the application of advanced probes and the exploration of non-equilibrium phenomena.
The pyrochlore iridates R$_2$Ir$_2$O$_7$ have emerged as a unique playground for exploring exotic quantum phenomena arising from the intricate interplay of strong spin-orbit coupling, electron correlations, and geometric frustration. While bulk crystals of these materials have revealed a rich landscape of correlated and topological states, recent breakthroughs in epitaxial thin-film synthesis and heterostructure engineering unlocked an entirely new dimension of discovery. This brief Perspective reviews recent advancements highlighting how new tuning knobs such as dimensional confinement, epitaxial strain, and interfacial coupling can be used to manipulate the delicate balance of competing interactions. We discuss several key discoveries enabled by this approach including the realization of the magnetic Weyl semimetal phase in (111) oriented films, strain-engineered magnetic multipolar orders, the emergence of a chiral spin liquid-like state in the quasi-2D limit, and the discovery of novel electronically anisotropic states at interfaces between pyrochlore iridates and other quantum materials, such as spin ice pyrochlores. These findings showcase that low-dimensional pyrochlore iridates provide ample opportunities for both theory and experiment to unravel, control and ultimately design novel quantum states of matter. We conclude by outlining key open questions and future directions ranging from the synthesis of new heterostructures to the application of advanced probes and the exploration of non-equilibrium phenomena.
\end{abstract}

\section{Introduction}
\begin{figure}
 \centering
        \includegraphics[width=0.75\textwidth]{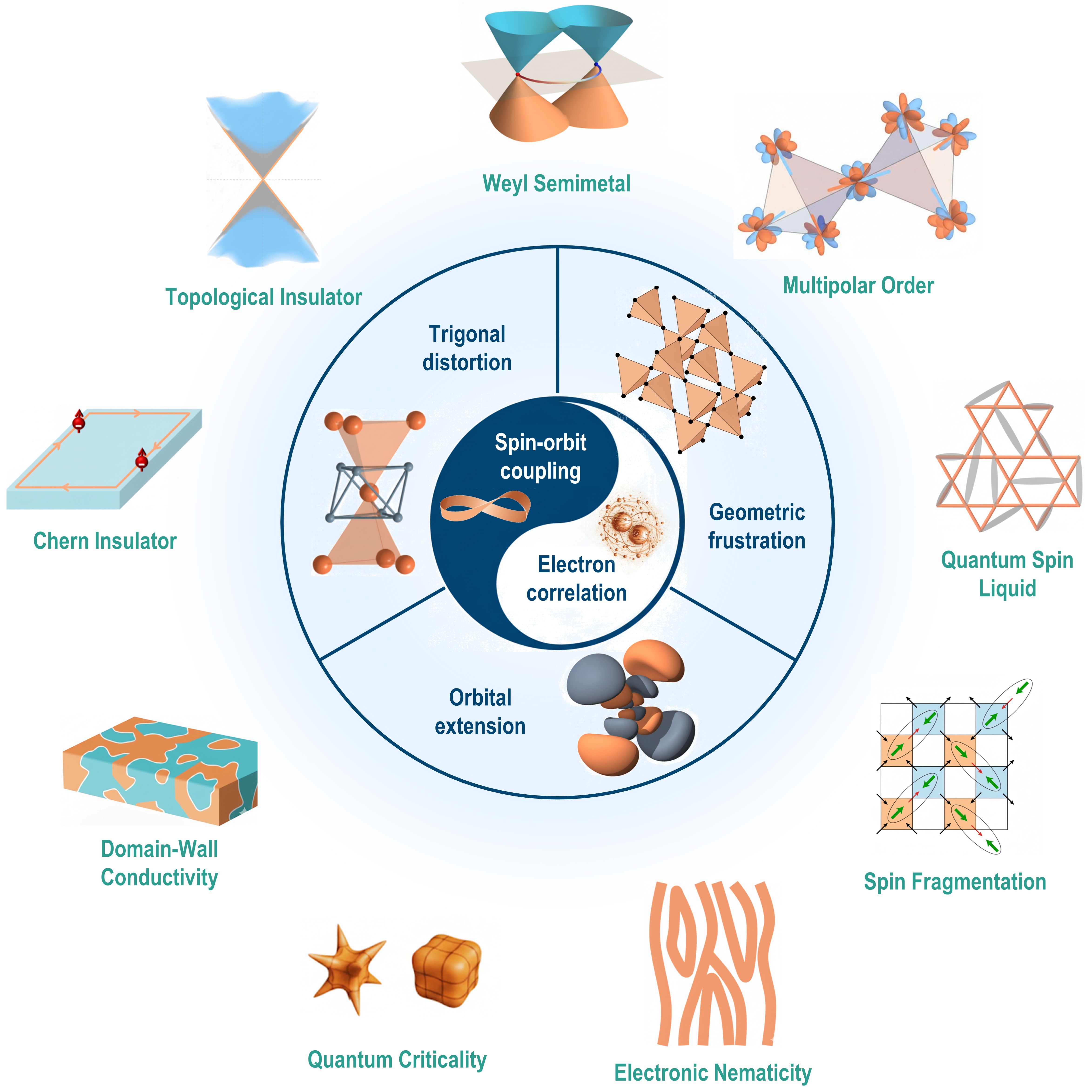}
 \caption{A schematic of the exotic quantum phenomena related to pyrochlore iridates. Driven by the synergy between strong spin-orbit coupling, electron-electron correlation, and three features (i.e., geometric frustration, orbital extension, local trigonal distortion) inherent to 5$d$ electrons on the corner-sharing tetrahedral network gives rise to a delicate balance between competing ground states. This results in the emergence of diverse novel and exotic quantum phases of matter, as illustrated.}
\label{fig1}
\end{figure}

Modern condensed matter physics has largely advanced along two major frontier trajectories: the continuing inquiry into electron-electron correlations linked to Mott transition and unconventional superconductivity among many emergent correlated states, and the exploration of strong spin-orbit coupling (SOC), the key ingredient for topological phases of matter~\cite{tokura-2022-natmater}. While these effects often individually dominate in various quantum material families, the current research has been focused on systems where these two interactions of different origins are of comparable strength, and once combined their interplay gives rise to entirely new physics~\cite{witczak-krempa-2014-annurev,schaffer2016recent}.
Although the role of SOC in driving topology in weakly interacting systems is well-established \cite{RevModPhys.83.1057, RevModPhys.82.3045}, its influence in the presence of strong electron correlations remains a profound and open question.
It is precisely in this regime that the 5$d$ pyrochlore iridates R$_2$Ir$_2$O$_7$ (R is a rare-earth element), have emerged as an exemplary platform~\cite{gardner-2010-rmp,Tokura_2025}. Here, correlation, electronic bandwidth, and SOC share comparable energy scales, offering a unique playground for emergent states where these fundamental interactions converge and compete in near perfect balance, as schematically outlined in Fig.~\ref{fig1}.

%This unique balance provides an ideal platform to investigate the rich variety of exotic electronic and magnetic states that arise from the synergy of these competing interactions, as schematically outlined in Fig.~\ref{fig1}.

\begin{figure}
 \centering
        \includegraphics[width=0.7\textwidth]{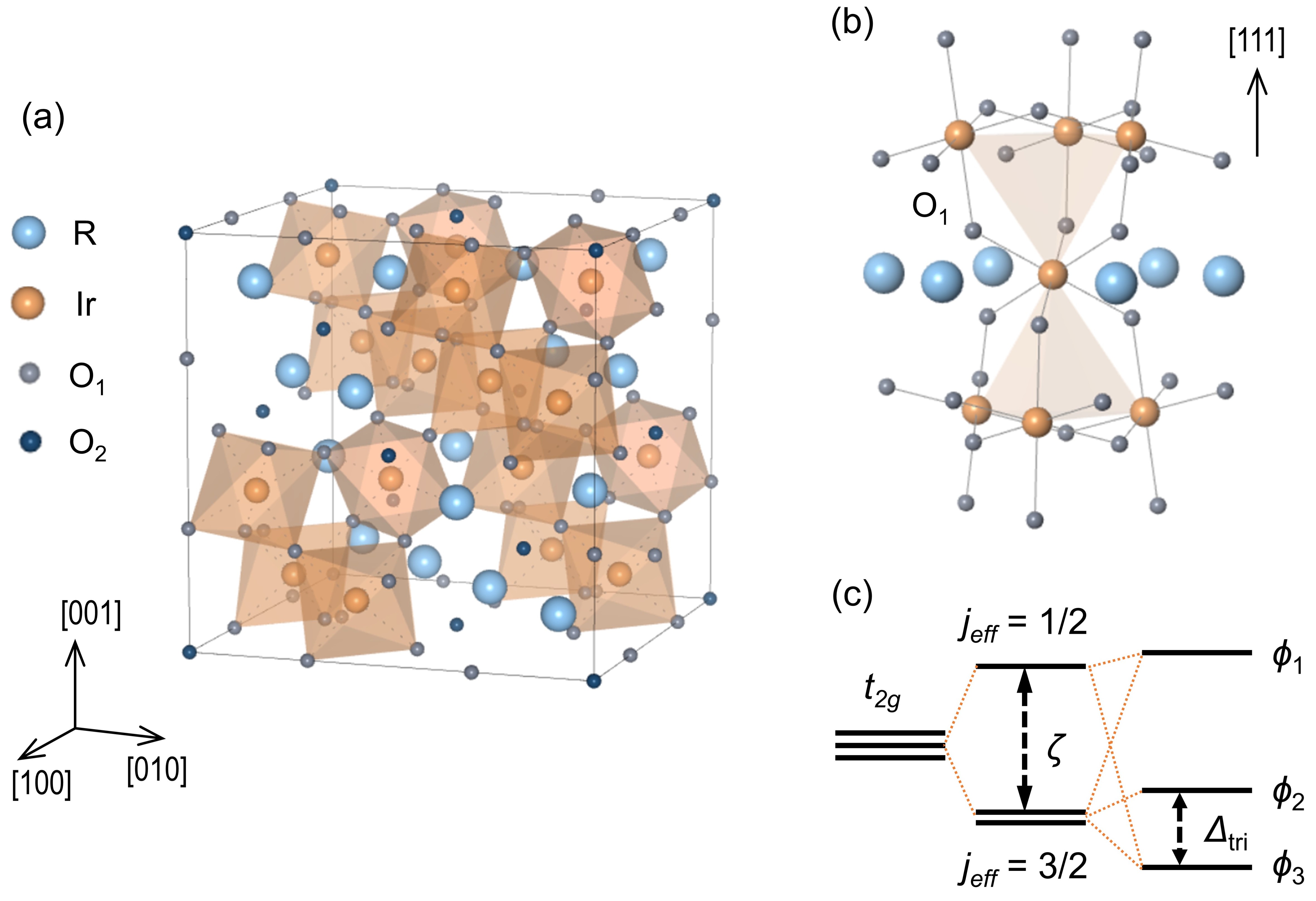}
 \caption{Crystal and electronic structure of pyrochlore iridates. (a) The conventional cubic unit cell of pyrochlore lattice showing the network of corner-sharing IrO$_6$ octahedra. The global three-dimensional connectivity is driven by the network of corner-sharing IrO$_6$ octahedra (orange), while the rare-earth R ions (blue) occupy the interstitial cavities forming a distorted scalenohedral coordination (R-O bonds are omitted for visual clarity). (b) A local view of two octahedra IrO$_6$ along the [111] axis. (c) Energy level diagram of Ir $t_{2g}$ orbitals. Strong spin-orbit coupling splits the $t_{2g}$ manifold into $j_{\text{eff}} = 3/2$ and $j_{\text{eff}} = 1/2$ levels. The trigonal crystal field $\Delta_{\text{tri}}$ further renders a hybridization among the $t_{2g}$ orbitals and leads to three discrete energy levels.}
\label{fig2}
\end{figure}

The pyrochlore iridates crystallize in the cubic \textit{Fd$\bar{3}$m} space group and can be viewed as a distorted $2\times2\times2$ superstructure of a cation-ordered yet anion-deficient fluorite lattice. The conventional cubic unit cell, approximately 10 $\textrm{\AA}$ in size, contains eight R$_2$Ir$_2$O$_7$ formula units (Fig.~\ref{fig2}a).  
The structure consists of two interpenetrating sublattices of corner-sharing tetrahedra: one formed by the rare-earth R ions at the 16c Wyckoff site and the other by the Ir ions at the 16d site. These two sublattices are displaced relative to each other by half a unit cell along the body diagonal. Within this framework, the R and Ir cations exhibit distinct oxygen coordination environments. 
Each R ion is surrounded by eight oxygen atoms located at both the 48f and 8a sites (denoted as O$_1$ and O$_2$), forming a distorted scalenohedral geometry, whereas the Ir ion is exclusively coordinated by six 48f oxygen atoms, forming the characteristic IrO$_{6}$ octahedra network shown in Fig.~\ref{fig2}(a).
The precise position of this 48f oxygen is crucial, as it dictates the degree of local trigonal distortion of each IrO$_{6}$ octahedron along the $\langle$111$\rangle$ direction (Fig.~\ref{fig2}(b)). This distortion modulates the crystal-field environment and can be systematically tuned through chemical substitution at the R site~\cite{witczak-krempa-2014-annurev,gardner-2010-rmp,Tokura_2025}. A smaller R$^{3+}$ ionic radius effectively exerts chemical pressure that shifts the 48f oxygen position, altering the Ir-O-Ir bond angles and enhancing the trigonal distortion. This structural control serves as a powerful handle for tuning the electronic and magnetic ground states of the pyrochlore iridates ~\cite{yang2010topological}.

Unlike their 3$d$ and 4$d$ counterparts where the on-site Coulomb repulsion $U$ typically dominates, the 5$d$ iridium ions possess spatially extended orbitals, leading to moderate-to-large electronic bandwidths ($W$) that effectively reduce $U$. 
Meanwhile, the large atomic number of iridium generates a strong SOC ($\zeta$), which splits the partially filled $t_{2g}$ electronic manifold into a fully occupied $j_{\text{eff}}$ = 3/2 quartet and a half-filled $j_{\text{eff}}$ = 1/2 doublet. The local trigonal distortion further introduces a crystal field splitting ($\Delta_{\text{tri}}$), lifting the degeneracy of the $j_{\text{eff}}=3/2$ states and produces three discrete energy levels within the $t_{2g}$ manifold~\cite{shinaoka-2019-jpcm} (see Fig.~\ref{fig2}(c)). 
The coexistence of comparable energy scales $U/W$, $\zeta$, and $\Delta_{\text{tri}}$ places pyrochlore iridates in a unique regime where electron correlations, SOC, and lattice effects compete yet cooperate. Moreover, this delicate balance makes pyrochlore iridates remarkably sensitive to external tuning parameters and in turn enables a range of quantum phase transitions. For instance, modest pressure~\cite{yang2010topological}, magnetic fields~\cite{tian2016field}, strain~\cite{kondo2015quadratic}, or chemical substitution~\cite{wan2011topological,zhang2017metal} can drive the system across a metal–insulator transition~\cite{pesin2010mott,witczak2010gauge}. Such high tunability places these materials in close proximity to exotic quantum critical points, where unconventional non-Fermi liquid behavior has been both theoretically proposed and experimentally observed~\cite{moon2013non,savary2014coulombic,yang2014quantum,shinaoka-2015-prl,boettcher2017anisotropy,moser2024quasiuniversality}.
Beyond these intrinsic interactions, the physics of pyrochlore iridates can be enriched by additional interactions such as the coupling between itinerant 5$d$ electrons and large localized 4$f$ moments on the R-site~\cite{chen2012magnetic,yao2018pr,ueda2022experimental,moser2024quasiuniversality} along with subdominant Hund’s exchange~\cite{georges2013strong}.

%The consequence is a system where $U$, $\zeta$, and $\Delta_{\text{tri}}$ are all crucial and comparable energy scales, allowing their cooperative interplay to makes the pyrochlore iridates remarkably sensitive to external stimuli, leading to a variety of quantum phase transitions. For example, the system can be driven across a metal-insulator transition~\cite{pesin2010mott,witczak2010gauge} by applying modest pressure~\cite{yang2010topological}, external magnetic fields~\cite{tian2016field}, strain~\cite{kondo2015quadratic}, or by chemical doping~\cite{wan2011topological,zhang2017metal}. This high degree of tunability also positions these materials in close proximity to exotic quantum critical points, where unconventional non-Fermi-liquid behaviors have been proposed and observed~\cite{moon2013non, savary2014coulombic,yang2014quantum,shinaoka-2015-prl,boettcher2017anisotropy,moser2024quasiuniversality}.
%Furthermore, the physics can be enriched by additional interactions, such as the coupling of the itinerant 5$d$ electrons to the localized 4$f$ moments on the R-site~\cite{chen2012magnetic,yao2018pr,ueda2022experimental,moser2024quasiuniversality} and subdominant Hund's coupling~\cite{georges2013strong}.

From the magnetism viewpoint, all bulk pyrochlore iridates except for Pr$_2$Ir$_2$O$_7$ undergo a magnetic phase transition below approximately 160 K~\cite{gardner-2010-rmp,Tokura_2025}. 
The interplay of correlations, SOC, and the inherent geometric frustration of the lattice conspires to stabilize a non-coplanar magnetic ground state, namely the `all-in-all-out' (AIAO, also termed as `4-in-0-out') antiferromagnetic (AFM) long-range order. This magnetic structure, which breaks time-reversal symmetry, serves as a theoretical foundation to achieving non-trivial band topology, such as the magnetic Weyl semimetal (WSM)~\cite{wan2011topological,yang2011quantum,witczak2012topological} or axion insulator~\cite{wan2011topological,go2012correlation}. However, it is important to point out this picture is still under active scrutiny~\cite{nakayama2016slater,wang2017electron,wang2017noncollinear}.
Experimentally, efforts to verify these topological phases have achieved significant and yet partial success. For example, the evidence for a WSM phase has emerged from the observations of a spontaneous anomalous Hall effect~\cite{fujita2015odd,ueda2018spontaneous,liu2021magnetic,lv2021experimental}, while the axion insulator which is anticipated to exhibit a quantized magnetoelectric response~\cite{tokura-2019-nrp}, remains elusive.
A striking counterexample arises in Pr$_2$Ir$_2$O$_7$, which notably avoids the long-range AIAO order. Instead, it hosts a magnetically disordered yet chiral ground state, characterized by the spin-ice `2-in-2-out' short-range correlations---possibly realizing a chiral spin-liquid-like phase~\cite{onoda2010quantum,onoda2011quantum,flint2013chiral,lee2013rkky,machida2007unconventional,machida2010time,ueda2012topological}. This interpretation is further supported by observations of a topological Hall effect~\cite{guo2020spontaneous,fukuda2022highly}. 

While the theoretical landscape of pyrochlore iridates is remarkably intriguing, experimental progress has been historically hampered by a significant materials science challenge: the difficulty in synthesizing large, high-quality bulk single crystals suitable for advanced characterization techniques.
Recent breakthroughs in epitaxial thin-film growth, particularly solid-phase epitaxy, have provided a powerful alternative route to realizing single-crystalline pyrochlore iridate epitaxial films~\cite{fujita2015odd}. Readers interested in details of these growth techniques can refer to several comprehensive reviews~\cite{chakhalian-2020-aplmater,kim-2022-aplmater,gutierrez2025epitaxial}. 
It is important to emphasize this progress is crucial because dimensionality offers a new avenue for control. Thin-film realizations bridge the crossover between two- and three-dimensional (3D) regimes, enabling emergent phenomena that go beyond those observed or predicted in bulk crystals. In particular, theories predict that such films can host giant and quantized anomalous Hall conductance~\cite{yang2014emergent, hwang2016theory,PhysRevLett.114.016806}. Controlling the position and separation of the Weyl nodes through thickness, strain, or stacking configuration enables tunability of system's topological properties. In reduced dimensions, this can give rise to magnetoelectric or topological responses that are weak or absent in the bulk, opening opportunities for probing correlation-driven topological states where electronic interactions and SOC are of comparable strength.
%
%In this Perspective, we review a few selected recent advancements in pyrochlore iridates, emphasizing the physics emerging from epitaxial thin films and heterostructures. 
%This approach provides access to additional tuning knobs such as strain, dimensional confinement, and interfacial coupling, which are not readily available in bulk materials. We discuss key experimental results where such tuning has been used to manipulate the electronic and magnetic ground states, leading to the discovery of new topological and correlated phenomena. These findings highlight the ample opportunities the thin-film platform provides for exploring emergent quantum phenomena in this material family. We conclude by outlining key open questions and promising future research directions.

In this Perspective, we review a few selected recent advancements in pyrochlore iridates, emphasizing the physics emerging from epitaxial thin films and heterostructures. This approach provides access to additional tuning knobs such as strain, dimensional confinement, and interfacial coupling, which are not readily available in bulk materials. 
Experimentally, these parameters are realized through substrate lattice mismatch (typically inducing epitaxial strains of less than 1\%), control of film thickness (with precision on the order of nanometers), and the fabrication of atomically coherent interfaces.
To showcase these tuning capabilities, our discussion will primarily focus on compounds with non-magnetic R-site ions (e.g., R = Eu, Y), magnetic rare-earth ions (e.g., R = Nd, Pr, Tb), and post-transition metals (e.g., Bi), as well as their heterostructures with classical spin ice materials (e.g., $Dy_2Ti_2O_7$).
Specifically, this Perspective is organized as follows: Section~\ref{sec:WSM} focuses on the realization and transport signatures of the magnetic WSM phase through (111) orientation. Section~\ref{sec:hidden} discusses how epitaxial strain can be harnessed to engineer and detect hidden multipolar magnetic orders. Section~\ref{sec:yio} explores the emergence of quantum-disordered, chiral spin-liquid-like states in the quasi-two-dimensional limit. Finally, Section~\ref{sec:interface} highlights novel electronic anisotropies driven by interfacial coupling in iridate-based heterostructures. These findings highlight the ample opportunities the thin-film platform provides for exploring emergent quantum phenomena in this material family. We conclude in Section~\ref{sec:summary} by outlining key open questions and promising future research directions.

\begin{figure}
 \centering
        \includegraphics[width=0.7\textwidth]{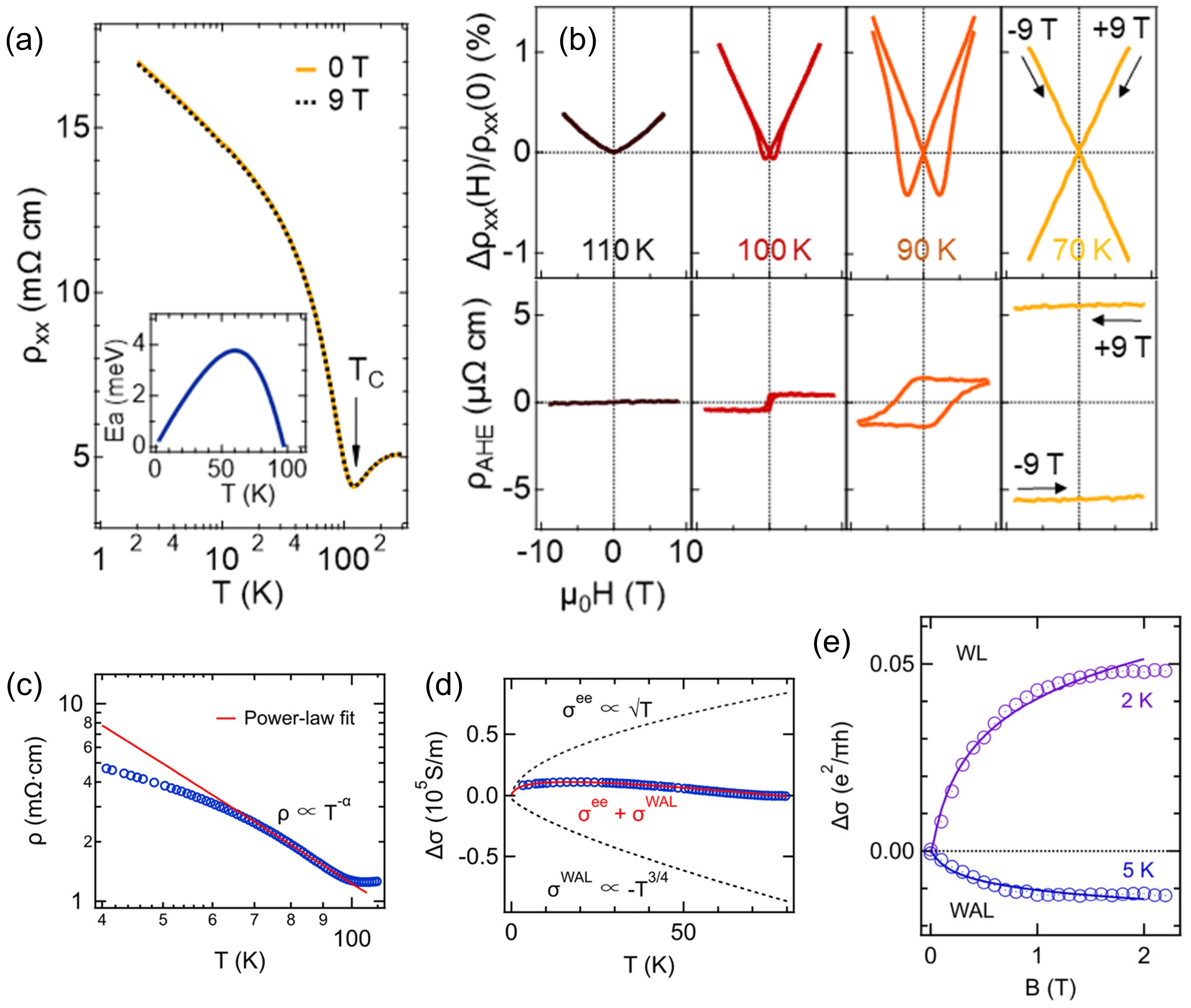}
 \caption{Transport signatures of WSM state in (111) Eu$_2$Ir$_2$O$_7$ thin films. (a) Temperature dependence of the longitudinal resistivity under 0 (orange) and 9 T (black) magnetic field, respectively. The onset of metal-semimetal transition is found at T$_C$ near 110 K. Inset: putative activation gap estimated using Arrhenius’s law. (b) Normalized longitudinal magnetoresistivity and transverse resistivity across T$_C$. The linear contribution from the ordinary Hall effect has been subtracted for
each transverse curve. (c) Power-law fit of resistivity data in the range of 70--100 K. (d) Low-temperature conductivity composed of two items, $\sigma^{\textit ee}$ from the electron–electron interaction (tendency of WL), and $\sigma^{\textit WAL}$ from the WAL effect. (e) WAL (blue) to WL (purple) crossover near the base temperature. Figures adapted from refs.~\cite{liu2021magnetic,wu-2024-apl}.}
\label{fig3}
\end{figure}

\section{Key Experimental Discoveries: From Bulk Puzzles to Thin-Film Engineering}\label{sec:exp}

\subsection{The Magnetic Weyl Semimetallic Phase}\label{sec:WSM}

An initial driver of research into pyrochlore iridates was the theoretical prediction that the non-coplanar AIAO AFM order would induce a WSM phase by breaking time-reversal symmetry~\cite{wan2011topological}. In this topological state, Weyl nodes---magnetic monopoles of Berry curvature in momentum space---are predicted to generate a large, intrinsic anomalous Hall effect (AHE). However, a major hurdle prevented the initial observation of this effect in bulk crystals: the high cubic symmetry of the pyrochlore lattice enforces a precise cancellation of the Berry curvature contributions from all Weyl nodes, leading to a vanishing net AHE~\cite{wan2011topological}.

The resolution to this symmetry problem emerged from thin-film engineering. It was proposed that fabricating films along the [111] direction would inherently break the requisite translational symmetries and unveil a net AHE~\cite{yang2014emergent}. This prediction spurred the successful growth of high-quality (111) epitaxial films of Eu$_{2}$Ir$_{2}$O$_{7}$, an ideal system due to its non-magnetic Eu$^{3+}$ ions which isolate the magnetism of the Ir sublattice. These films enabled the definitive observation of an intrinsic AHE emerging concurrently with the AIAO magnetic order~\cite{liu2021magnetic}. The AHE was remarkable, exhibiting a colossal coercive field while stemming from a state with a vanishingly small net magnetization, which serves as a key macroscopic transport signature of its topological origin (Fig.~\ref{fig3}(a)-(b)). This picture was further enriched by the discovery of a large topological Hall effect (THE) in these films, which might originate from the real-space topology of the non-coplanar spin texture~\cite{ghosh-2022-prb}.

Beyond this intrinsic Hall response, the WSM state also imparts more subtle quantum transport signatures. A key example is the weak antilocalization (WAL) effect, a quantum interference effect where the nontrivial Berry phase of Weyl fermions suppresses backscattering and increases conductivity. Recent magnetotransport studies on (111) Eu$_2$Ir$_2$O$_7$ films have successfully uncovered this signature that was previously `buried' by the dominant magnetic scattering effects \cite{wu-2024-apl}. Remarkably, these experiments also revealed a crossover from WAL to weak localization (WL) below 5 K (Fig.~\ref{fig3}(c)-(e)). This switching behavior is attributed to the growing influence of many-body interactions, providing a beautiful experimental window into the direct competition between band topology favoring WAL and electron correlations driving WL in an interacting WSM \cite{lu-2015-prb}. 

Recently, the elusive AFM domains in the WSM state were directly visualized for the first time by optical circular dichroism (CD) microscopy and Kerr effect on (111) Eu$_2$Ir$_2$O$_7$ thin films. It resolved the AIAO and AOAI domain configurations, confirming the presence of a ferroic magnetic octupole order. Critically, a large CD signal was detected at zero field, a striking observation for materials with no net magnetic moment. This effect is attributed to the topological Berry curvature from the Weyl nodes, highlighting how non-trivial band topology can manifest as strong magneto-optical responses even in compensated magnetic systems~\cite{han-2025-arxiv}.

Epitaxial strain has been deployed as another powerful knob to realize the WSM phase. In bulk form, Pr$_{2}$Ir$_{2}$O$_{7}$ is a paramagnetic Luttinger semimetal, distinguished by a quadratic band touching at the Fermi level~\cite{kondo2015quadratic}. It was demonstrated that a small compressive strain ($\approx$~0.2\%) applied to (111) thin films could tip the balance of magnetic interactions to stabilize the AIAO AFM order absent in its bulk, and drive the system from a Luttinger semimetal into a magnetic WSM~\cite{li-2021-advmater}. 
The resulting strain-induced state exhibited distinct transport signatures such as the AHE and a negative longitudinal magnetoresistance (a hallmark of the chiral anomaly), revealing compelling evidences for an interacting WSM state~\cite{guo2020spontaneous,li-2021-advmater,ohtsuki-2019-pnas}.

\subsection{Strain Engineering of Hidden Magnetic Orders}\label{sec:hidden}

\begin{figure}
 \centering
        \includegraphics[width=0.7\textwidth]{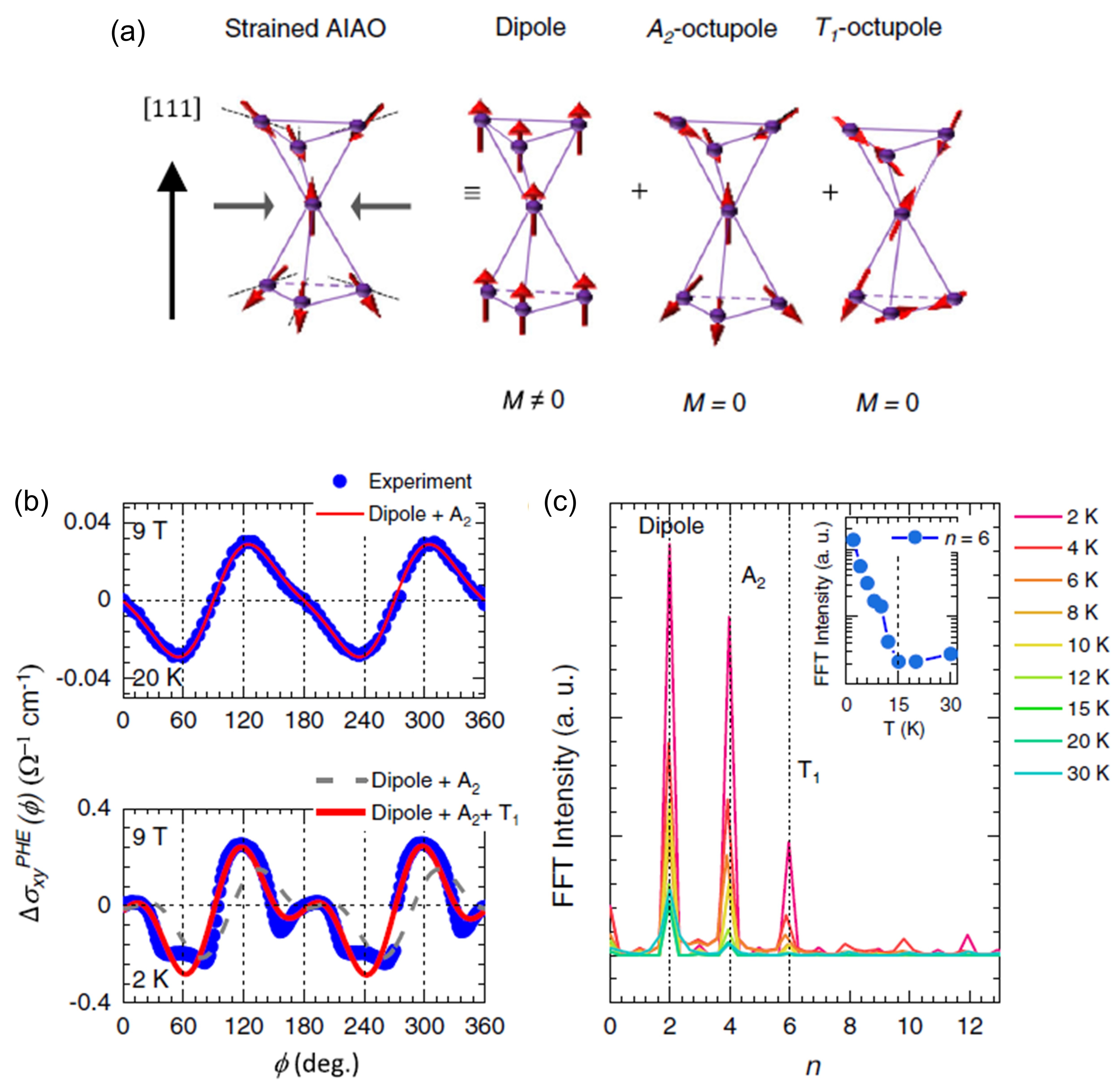}
 \caption{Probing cluster magnetic multipoles in strained Nd$_{2}$lr$_{2}$O$_{7}$ using the Planar Hall Effect (PHE).(a) Schematic illustrating how compressive strain along the [111] direction modifies the all-in-all-out (AIAO) magnetic configuration. This strained structure can be decomposed into three distinct cluster magnetic multipoles: a magnetic dipole ($M \neq 0$), an $A_{2}$-octupole ($M=0$), and a $T_{1}$-octupole ($M=0$)~\cite{kim-2020-sciadv}. (b) Angle-dependent PHE conductivity, $\Delta\sigma_{xy}^{PHE}(\phi)$, measured at an external field of 9 T. At 20 K (top), the oscillation is well-described by contributions from the dipole and the $A_{2}$-octupole. At 2 K (bottom), a complex oscillation appears that cannot be explained by the dipole and $A_{2}$-octupole alone (dashed gray line). An accurate fit (solid red line) requires the inclusion of the $T_{1}$-octupole's contribution. (c) Fast Fourier Transform (FFT) intensity of the PHE oscillations across different temperatures. The primary plot reveals distinct second ($n=2$), fourth ($n=4$), and sixth ($n=6$) harmonics, which are the unique signatures of the dipole, $A_{2}$-octupole, and $T_{1}$-octupole, respectively. The inset shows the temperature dependence of the sixth harmonic's intensity, confirming that the $T_{1}$-octupole order emerges below 15 K. Figure adapted from ref.~\cite{song-2022-natcommun}. Note, readers interested in the explicit microscopic Hamiltonian for the electronic hopping and magnetic exchange terms under strain are referred to the detailed theoretical derivations in refs.~\cite{kim-2020-sciadv,song-2022-natcommun}.}
\label{fig4}
\end{figure}

Beyond stabilizing known ground states, epitaxial strain serves as a powerful means to engineer new magnetic and topological phases in pyrochlore iridates. Its effectiveness arises from the direct modification of the Ir-O-Ir bond angles and Ir-Ir distances, which tunes the electronic hopping parameters and shifts the delicate balance among competing magnetic exchange interactions.

%A particularly notable example is demonstrated by straining Nd$_{2}$Ir$_{2}$O$_{7}$ thin films, where the disruption of the AIAO order exposes the presence of `hidden' multipolar orders. Specifically, while conventional magnetism is described by a dipole moment, complex AFM states on specific lattices can possess higher-rank magnetic multipoles that are undetectable by bulk magnetization probes. In compressively strained (111) Nd$_{2}$Ir$_{2}$O$_{7}$ films, it was predicted that the strain-distorted AIAO spin structure could be decomposed into not only the conventional $A_2$-octupole but also an emergent $T_1$-octupole moment~\cite{kim-2020-sciadv}. While the $A_2$-octupole preserves the symmetries that cancel the AHE, the strain-induced $T_1$-octupole breaks them yielding a large AHE even in the absence of net magnetization.

A particularly notable example is demonstrated by straining Nd$_2$Ir$_2$O$_7$ thin films, where the disruption of the AIAO order exposes the presence of `hidden' multipolar orders~\cite{kim-2020-sciadv,PhysRevB.98.125103}. Specifically, while conventional magnetism is described by a dipole moment, complex AFM states on specific lattices can possess higher-rank magnetic multipoles that are undetectable by bulk magnetization probes. In the bulk $O_h$ cubic symmetry, the macroscopic anomalous transport responses of the AIAO state strictly vanish. However, applying compressive strain along the [111] direction lowers the crystal symmetry to the $D_{3d}$ point group. This structural distortion alters the Ir-O-Ir bond angles, modulating the electronic hopping amplitudes and shifting the delicate balance among competing magnetic exchange interactions. Consequently, the strain-distorted AIAO spin structure can be decomposed into not only the conventional $A_2$-octupole but also an emergent $T_1$-octupole moment (Fig.~\ref{fig4}(a)). While the $A_2$-octupole preserves the symmetries that cancel the AHE, the strain-induced $T_1$-octupole breaks them, yielding a large AHE even in the absence of net magnetization.

%In addition, the complex magnetic order could be probed by the planar Hall effect (PHE). Based on the symmetry argument, it was demonstrated that different multipoles can generate angular patterns of distinct symmetries in the PHE signal. For instance, the dipole, $A_2$-octupole, and $T_1$-octupole produce second, fourth, and sixth harmonics, respectively. By measuring the higher harmonics, one could identify the presence and temperature dependence of each specific multipolar contribution to the hidden order from an electrical magneto-transport fingerprinting technique~\cite{song-2022-natcommun}.

In addition, these complex magnetic multipoles can be directly probed by the planar Hall effect (PHE)~\cite{song-2022-natcommun,PhysRevLett.130.266703}. Based on symmetry arguments, different multipoles generate angular patterns of distinct symmetries in the macroscopic PHE signal. When an in-plane magnetic field is rotated, the PHE resistance oscillates. By decomposing these oscillations into a Fourier series, the twofold (second harmonic), fourfold (fourth harmonic), and sixfold (sixth harmonic) angular dependencies are revealed as the direct macroscopic fingerprints of the magnetic dipole, the $A_2$-octupole, and the $T_1$-octupole, respectively, providing a direct electrical readout of these hidden orders (Fig.~\ref{fig4}(b)-(c)). By measuring these higher harmonics, one can precisely identify the presence and temperature dependence of each specific multipolar contribution to the hidden order utilizing an electrical magneto-transport fingerprinting technique.

\subsection{Disordered States and Emergent Spin Liquid Physics}\label{sec:yio}

\begin{figure}
\centering
	\includegraphics[width=0.7\textwidth]{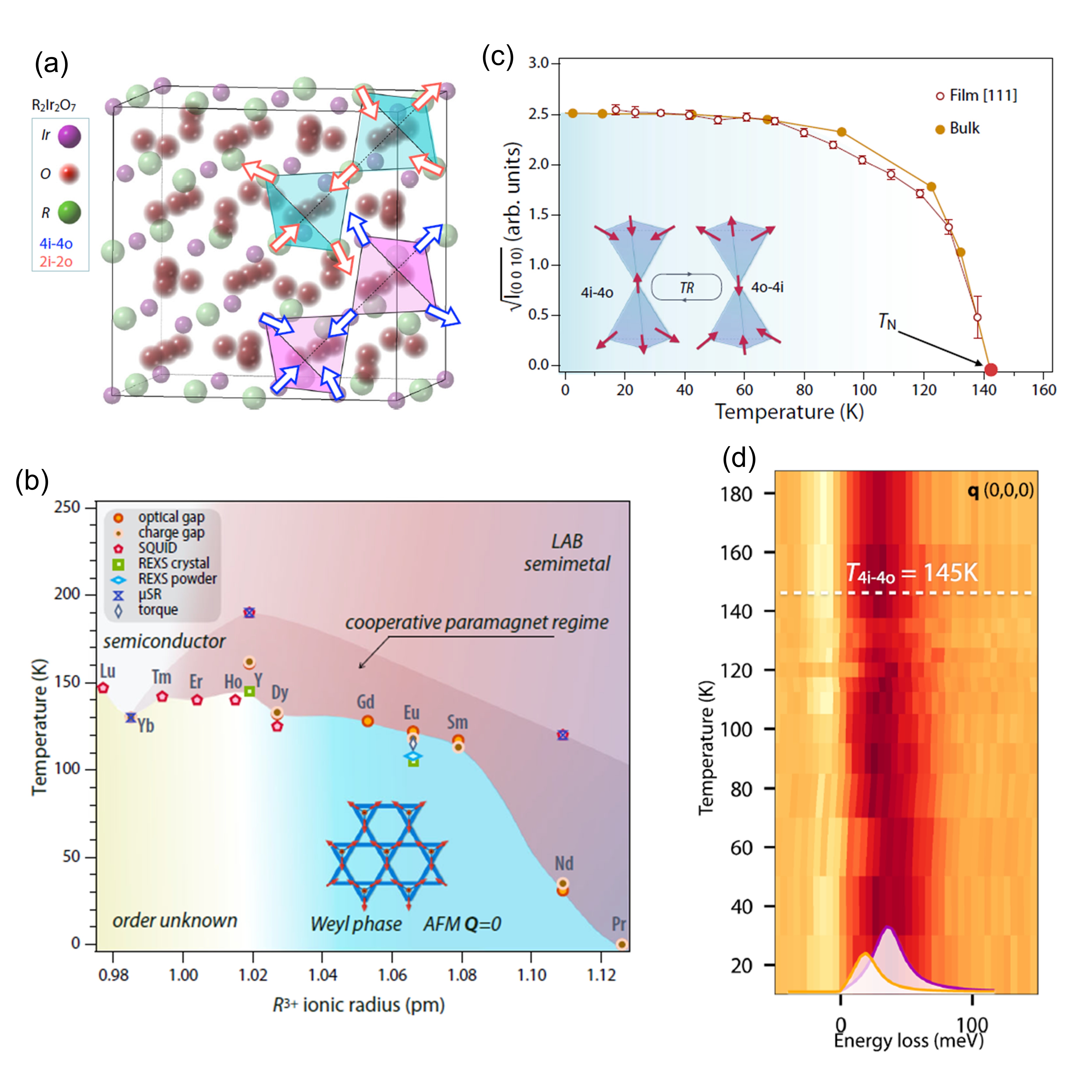}
\caption{Probing magnetic ground state and excitations in Y$_{2}$lr$_{2}$O$_{7}$ with x-ray scattering.(a) The pyrochlore lattice features two interpenetrating sublattices of corner-sharing tetrahedra. In the ordered ground state, possible spin configurations include ``two-in-two-out'' (red arrows), as in the titanates, or ``four-in-four-out'' (blue arrows), as in the iridates. (b) Experimental phase diagram of the pyrochlore iridates, featuring optical and charge gap \cite{Ueda2016}, SQUID \cite{Lefrancois2017, Cathelin2020, Vlaskova2020, Klicpera2020}, REXS \cite{Tomiyasu2011, Donnerer2016, liu2021magnetic}, and torque magnetometry \cite{Liang2017} data. Spectroscopic studies have been limited for smaller A-site radii due to the difficulty of growing large-size single crystals. $\mu$SR data indicates the presence of a quantum cooperative paramagnet regime for Y$_2$Ir$_2$O$_7$ and Nd$_2$Ir$_2$O$_7$ \cite{Disseler2012_NIO, Disseler2012a}. (c) Temperature-dependent scans around the (0 0 10) were taken, and the peak intensity was plotted as a function of temperature (open circles). The results have a similar profile to the muon precession frequency of Y$_{2}$lr$_{2}$O$_{7}$ (solid circles) \cite{Disseler2012a}. Inset shows the time-reversal pair of the 4i-4o and 4o-4i ordering. (d) Color map of temperature-dependent RIXS data after subtraction of the elastic line and magnetic background. A magnon gap is clearly seen, indicating the presence of gapped excitations that persist above $T_N$. The fitted magnon modes for $T=10\,$K are plotted at the bottom of the figure. Figure adapted from ref.~\cite{Terilli2025}.}
\label{fig5} 
\end{figure}

Pyrochlore iridates have also emerged as an important system for investigating quantum disordered states, where strong quantum fluctuations can destabilize long-range magnetic order. A compelling example is Y$_{2}$Ir$_{2}$O$_{7}$, a system with non-magnetic Y$^{3+}$ ions that keeps the Ir magnetism unperturbed. Recent experiments on high-quality Y$_{2}$Ir$_{2}$O$_{7}$ films revealed a fascinating evolution of its magnetic behavior, transitioning from a bulk-like phase characterized by unusually robust short-range correlations to a fully realized quantum disordered state in the quasi-two-dimensional limit~\cite{Terilli2025,liu-2024-natcommun}.

More specifically, advanced scattering measurements on thick Y$_{2}$Ir$_{2}$O$_{7}$ films uncovered an unexpected magnetic behavior. %Resonant elastic x-ray scattering (REXS) \cite{Sagayama2013, Tomiyasu2011, Donnerer2016, Chun2018}confirmed a second-order phase transition into the AIAO state with a N\'{e}el temperature of 145~K and a critical exponent ($\beta$) of 0.30$\pm$0.04, consistent with the 3D Ising-like magnetic order as observed in other members of R$_2$Ir$_2$O$_7$~\cite{Terilli2025, Odor2004,Sagayama2013, Tomiyasu2011, Donnerer2016, Chun2018}. However, resonant inelastic x-ray scattering (RIXS) revealed that while the system develops a clear magnon gap in the ordered state~\cite{Terilli2025}, these gapped magnetic excitations abnormally persist up to 175~K, well above the ordering temperature~\cite{Terilli2025}. This contrasts sharply with other iridates like Eu$_{2}$Ir$_{2}$O$_{7}$, where the magnon gap closes as the temperature approaches $T_N$~\cite{Chun2018}. This unusual persistence of magnetic correlations in Y$_{2}$Ir$_{2}$O$_{7}$ is likely supported by the absence of a concomitant structural or sharp metal-insulator transition at $T_N$~\cite{Liu2020YIO}, which is present in Eu$_{2}$Ir$_{2}$O$_{7}$~\cite{liu2021magnetic, Das_2022}, pointing to a ``cooperative paramagnet'' regime above $T_N$, where robust short-range correlations survive the loss of long-range order.
Resonant elastic x-ray scattering (REXS)~\cite{Sagayama2013, Tomiyasu2011, Donnerer2016, donnerer-2019-jpcm, Chun2018} confirmed a second-order phase transition into the AIAO state with a N\'{e}el temperature of 145 K and the 3D Ising-like magnetic critical exponent $\beta$ of 0.30$\pm$0.04~\cite{Terilli2025, Odor2004,Sagayama2013, Tomiyasu2011, Donnerer2016, Chun2018}. However, resonant inelastic x-ray scattering (RIXS) revealed that while the system develops a clear magnon gap in the ordered state~\cite{Terilli2025}, these gapped magnetic excitations abnormally persist up to 175~K, well above the ordering temperature. This contrasts sharply with other iridates like Eu$_{2}$Ir$_{2}$O$_{7}$, where the magnon gap closes as the temperature approaches $T_N$~\cite{Chun2018}. This surprising persistence of magnetic correlations in Y$_{2}$Ir$_{2}$O$_{7}$ is likely supported by the absence of a concomitant structural or sharp metal-insulator transition at $T_N$~\cite{Liu2020YIO}, which is present in Eu$_{2}$Ir$_{2}$O$_{7}$~\cite{liu2021magnetic, Das_2022}, pointing towards a `cooperative paramagnet' regime emerging above $T_N$, where robust short-range correlations survive the loss of long-range order, as captured by the anomalous persistence of gapped magnetic excitations in Fig.~\ref{fig5}.

\begin{figure}
\centering
	\includegraphics[width=0.65\textwidth]{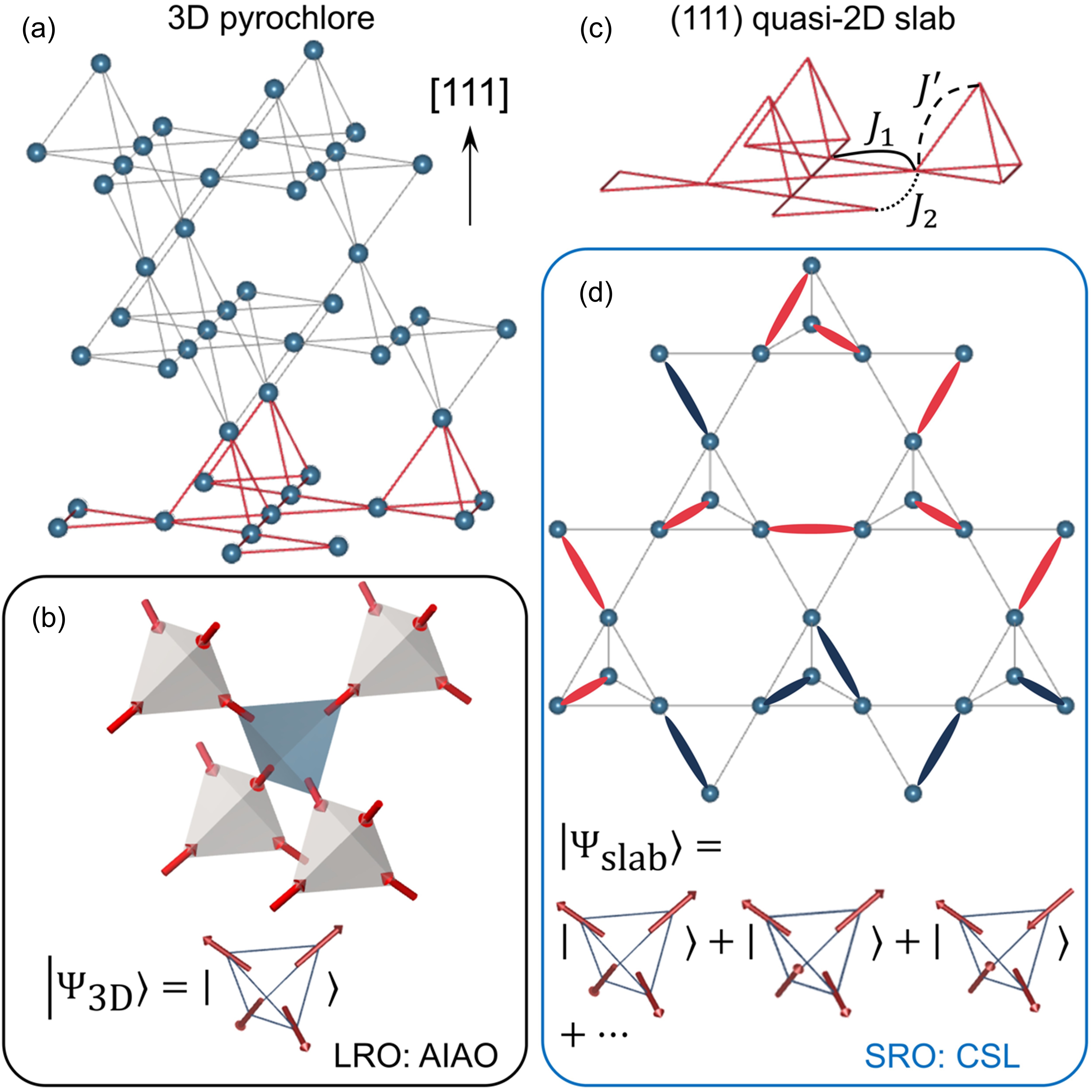}
\caption{Comparison of magnetic ground states in bulk (3D) and thin-film (quasi-2D) pyrochlore iridates. 
(a) The 3D pyrochlore lattice is constructed from Ir atoms arranged in a network of corner-sharing tetrahedra. 
(b) In the bulk material, the magnetic ground state is the ``all-in-all-out (AIAO)'' antiferromagnetic spin configuration, which exhibits Long-Range Order (LRO).
(c) In a quasi-2D (111) thin film, the lattice is transformed into a slab composed of alternating kagome and triangle atomic layers, where various magnetic exchange interactions ($J_1$, $J_2$, $J'$) compete.
(d) This dimensional confinement suppresses long-range ordering and stabilizes a chiral spin-liquid-like (CSL) state with only short-range order (SRO). This quantum disordered state is conceptualized as a superposition of fluctuating spin dimers covering the lattice. Figure adapted from ref.~\cite{liu-2024-natcommun}.}
\label{fig6}
\end{figure}

This tendency towards disorder becomes dramatically enhanced by dimensional confinement. In (111)-oriented Y$_{2}$Ir$_{2}$O$_{7}$ thin films, a clear thickness-dependent crossover was recently uncovered: while thicker 100~nm films demonstrated the expected AIAO long-range order, the order was completely suppressed in thinner set of $\le$ 30 nm films~\cite{liu-2024-natcommun}. In this quasi-2D limit, a new quantum disordered phase emerged that still could break time-reversal symmetry, as evidenced by a prominent AHE. Further RIXS measurements revealed the presence of dispersionless magnetic excitations---a hallmark of localized, non-propagating modes. The combination of the AHE signal in the absence of static magnetic order and the presence of dispersionless magnetic excitations suggests a chiral spin-liquid-like state, i.e., an entangled magnetic phase without static order~\cite{liu-2024-natcommun}. This progression illustrates how dimensionality can be used as a tuning parameter to amplify quantum fluctuations and unveil hidden and potentially highly entangled ground states, schematically summarized in Fig.~\ref{fig6}.

The unusual character of these excitations, however, poses a theoretical challenge. Conventional spin-wave analysis, typically used to treat magnon modes in pyrochlore iridates, failed to capture the behavior of the higher-energy mode near the $\Gamma$-point in Y$_{2}$Ir$_{2}$O$_{7}$. Similar discrepancies have been reported in nonstoichiometric Tb$_{2}$Ir$_{2}$O$_{7}$~\cite{faure2024}, suggesting that a random-phase approximation approach may be better suited for the smaller rare-earth site compounds~\cite{Lee2013}. Alternatively, recent studies suggest that incorporating quantum fluctuations within a nonlinear spin-wave theory framework can offer a more consistent description of these magnetic excitations~\cite{Wang_2025}.

\subsection{Novel Interfacial Phenomena}\label{sec:interface}

\begin{figure}
\centering
	\includegraphics[width=0.7\textwidth]{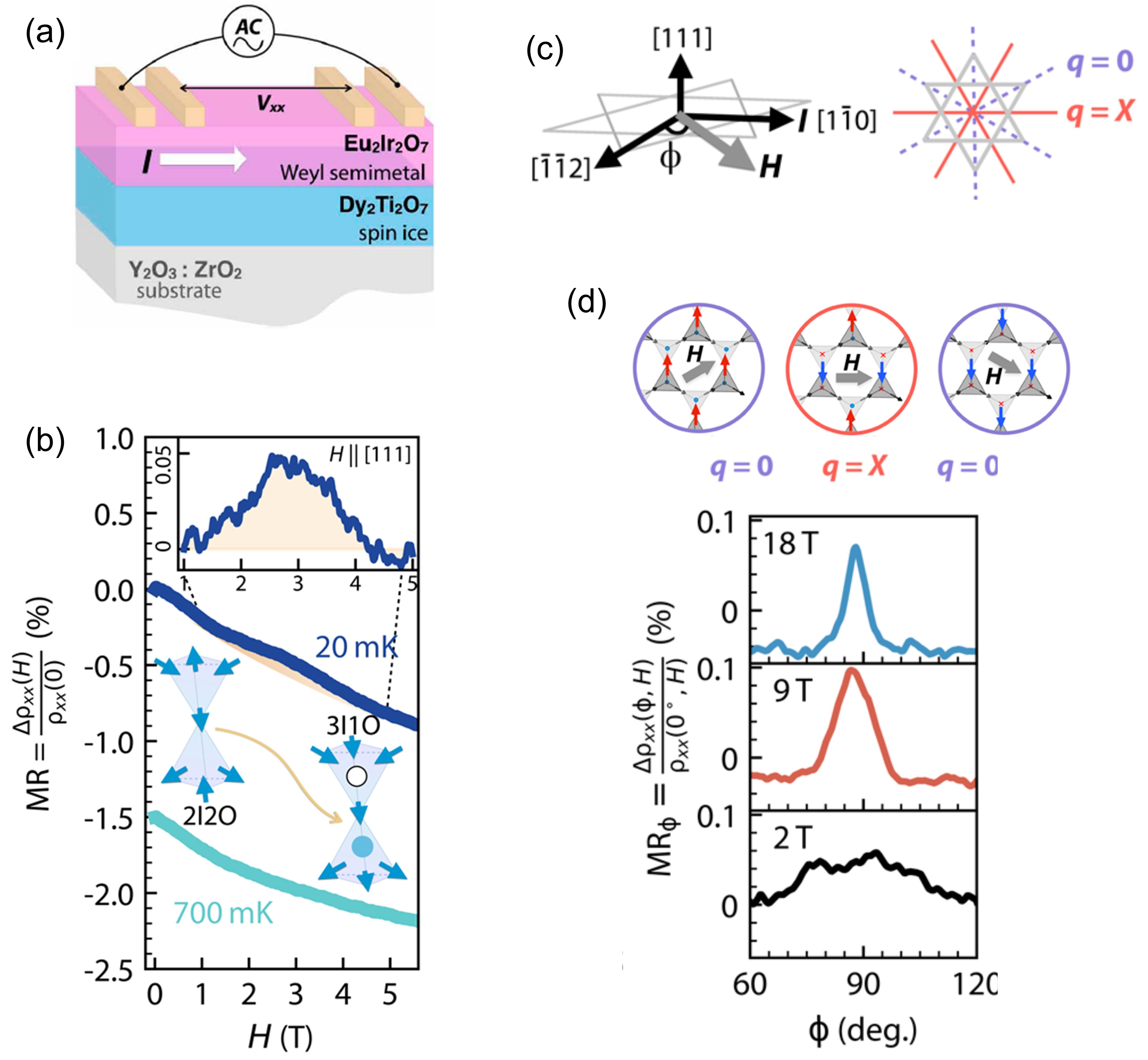}
\caption{ 
Emergent electronic anisotropy at a Weyl semimetal/spin ice interface probed by magnetotransport. 
(a) Schematic of the four-probe magnetotransport measurement setup on the Eu$_2$lr$_2$O$_7$/Dy$_2$Ti$_2$O$_7$ (EIO/DTO) heterostructure.
(b) Magnetoresistance (MR) with the magnetic field applied perpendicular to the interface ($H \parallel [111]$). At 20 mK, an anomalous bump feature appears, corresponding to the field-induced transition in the DTO layer from a two-in-two-out (2I2O) spin ice state to a three-in-one-out (3I1O) magnetic monopole state. 
(c) Geometry for in-plane field rotation, where the angle $\phi$ is swept. The DTO layer is driven between $q=0$ and $q=X$ magnetic phases, which occur at $60^{\circ}$ and $30^{\circ}$ intervals, respectively.
(d) In-plane angular MR ($MR_{\phi}$) reveals a field-dependent rotational symmetry. At lower magnetic fields (2 T), the response shows a sixfold anisotropy as expected from the symmetry of the underlying lattice. At high magnetic fields (9 T and 18 T), twofold anisotropy emerges and becomes dominant with resistance maxima occurring when the DTO layer is in the $q=X$ phase, signaling a symmetry-broken many-body state at the interface. Figure adapted from ref.~\cite{wu-2025-sciadv}.
}
\label{fig7}
\end{figure}

Engineering interfaces between pyrochlore iridates and other quantum materials opens another powerful avenue for discovering and controlling emergent phenomena. By placing a pyrochlore iridate in contact with a different material, novel electronic and magnetic states can arise at the two-dimensional boundary that do not exist in either constituent material alone.

Interfaces between magnetic pyrochlores remain a nascent field of research. An early demonstration of this approach was the controlled creation of a single magnetic domain wall at the heterointerface constructed by pyrochlore iridates. In bulk R$_2$Ir$_2$O$_7$, numerous randomly oriented magnetic domains can coexist, but in a Eu$_{2}$Ir$_{2}$O$_{7}$/Tb$_{2}$Ir$_{2}$O$_{7}$ heterostructure, a well-defined planar interface was engineered between an AIAO domain and its time-reversed counterpart, the AOAI domain. Transport measurements across this interface revealed a distinct metallic conduction channel accompanied by ferroic hysteresis. This result confirmed the theoretical prediction that such topological domain walls can host mobile metallic states~\cite{fujita-2016-prb}.

%An early example of this approach was the creation of a single magnetic domain wall at the heterointerface constructed by pyrochlore iridates. While bulk R$_2$Ir$_2$O$_7$ contain many randomly oriented magnetic domains, a heterostructure of Eu$_{2}$Ir$_{2}$O$_{7}$/Tb$_{2}$Ir$_{2}$O$_{7}$ was used to controllably create a single, planar interface between an AIAO domain and its time-reversed partner, an all-out-all-in (AOAI) domain. Transport measurements on this interface revealed a distinct metallic conduction channel with ferroic hysteresis, confirming theoretical predictions that these topological domain walls host mobile metallic states~\cite{fujita-2016-prb}.

More recently, interfaces between a Weyl semimetal Eu$_{2}$Ir$_{2}$O$_{7}$ and a classical spin ice Dy$_{2}$Ti$_{2}$O$_{7}$ have been fabricated~\cite{wu-2025-sciadv}. Spin ice is a frustrated magnetic state following the local ``2-in-2-out'' ice rules akin to proton disorder in water ice. These heterostructures exhibited a striking symmetry-broken six-fold anisotropic angular magnetoresistance at sub-Kelvin temperatures. This novel electronic anisotropy was not a property of the Weyl semimetal Eu$_{2}$Ir$_{2}$O$_{7}$ film itself but was directly inherited from the underlying magnetic spin texture of the insulating spin ice. When the magnetic field exceeds 5 T, the system transitions to a two-fold symmetric state, indicative of a field-induced spin-nematic phase~\cite{PhysRevB.109.014438}.  The effect was traced to an interfacial Kondo-like coupling, where the itinerant Weyl fermions located in the Fermi arcs of the iridate scatter off the magnetic Dy moments, effectively `reading out' the state of the underlying frustrated magnet~\cite{wu-2025-sciadv}. This work demonstrates a powerful general principle: a metallic quantum structure can be used as an extremely sensitive electronic probe of the exotic correlated states present in an adjacent insulator, evidenced by the emergent interfacial rotational symmetry breaking shown in Fig.~\ref{fig7}. 

%More recently, interfaces between a Weyl semimetal Eu$_{2}$Ir$_{2}$O$_{7}$ and a classical spin ice Dy$_{2}$Ti$_{2}$O$_{7}$ have been fabricated~\cite{wu-2025-sciadv}. Spin ice is a frustrated magnetic state following the local ``2-in-2-out'' spin rules, analogous to water ice. These heterostructures exhibited a striking six-fold anisotropic magnetoresistance at sub-Kelvin temperatures. This novel electronic anisotropy was not a property of the Weyl semimetal film itself but was directly inherited from the underlying magnetic state of the insulating spin ice. The effect was traced to an interfacial Kondo-like coupling, where the itinerant Weyl fermions of the iridate scatter off the magnetic Dy moments, effectively ``reading out'' the state of the frustrated magnet~\cite{wu-2025-sciadv}. This work demonstrates a powerful general principle: a metallic quantum material can be used as a sensitive electronic probe of the exotic correlated states present in an adjacent insulator. Engineering interfaces between pyrochlore iridates and other quantum materials opens another powerful avenue for discovering and controlling emergent phenomena. By placing a pyrochlore iridate in contact with a different material, novel electronic and magnetic states can be induced at the two-dimensional boundary that do not exist in either constituent material alone.

A different approach was also explored at interfaces where the metallic iridate is paramagnetic. In a heterostructure composed of a metallic Bi$_2$Ir$_2$O$_7$ film on a Dy$_2$Ti$_2$O$_7$ spin ice insulator, a novel anomalous magnetoresistance was discovered~\cite{zhang-2023-natcommun}. This transport signature, which emerged only at the interface, was shown to be a direct probe of the field-induced `kagome ice' state in the underlying insulator, where the spin ice ``2-in-2-out'' rule is broken. This work further established interfacial magnetotransport as a highly sensitive tool to detect exotic magnetic correlations and phase transitions in an adjacent insulating layer.

To integrate the diverse experimental advancements discussed above, Table~\ref{summary} provides a concise overview of representative thin-film pyrochlore iridate systems, highlighting how specific tuning parameters unlock a rich landscape of emergent quantum phenomena.

\begin{table}[htbp]
\centering
\caption{Summary of representative emergent phenomena in pyrochlore iridates thin films and heterostructures.}
\label{summary}
\renewcommand{\arraystretch}{1.3}
\begin{tabularx}{\textwidth}{llX}
\hline\hline
\textbf{System} & \textbf{Tuning Knob} & \textbf{Emergent Physics} \\
\hline
Eu$_2$Ir$_2$O$_7$ & Orientation & Magnetic WSM and intrinsic AHE \cite{fujita2015odd,liu2021magnetic} \newline WAL to WL crossover \cite{wu-2024-apl} \newline THE \cite{ghosh-2022-prb} \newline Optical CD of octupole domains \cite{han-2025-arxiv} \\
Y$_2$Ir$_2$O$_7$ & Orientation & Anomalous persistence of magnon gap and cooperative paramagnetism above $T_N$ \cite{Terilli2025} \\
 & Confinement & Chiral spin-liquid-like state in quasi-2D limit \cite{liu-2024-natcommun} \\
\hline
Pr$_2$Ir$_2$O$_7$ & Strain & Strain-induced AHE and interacting WSM \cite{guo2020spontaneous,li-2021-advmater,ohtsuki-2019-pnas} \\
Nd$_2$Ir$_2$O$_7$ & Strain & Strain-engineered $T_1$ multipolar order \cite{kim-2020-sciadv,PhysRevB.98.125103} \newline PHE harmonics from magnetic multipoles \cite{song-2022-natcommun}  \\
\hline
Eu$_2$Ir$_2$O$_7$/Tb$_2$Ir$_2$O$_7$ & Interface & Mobile metallic domain wall conduction \cite{fujita-2016-prb} \\
Eu$_2$Ir$_2$O$_7$/Dy$_2$Ti$_2$O$_7$ & Interface & Anisotropic MR at WSM/spin-ice interface \cite{wu-2025-sciadv} \\
Bi$_2$Ir$_2$O$_7$/Dy$_2$Ti$_2$O$_7$ & Interface & Anomalous MR probing `kagome ice' \cite{zhang-2023-natcommun} \\
\hline\hline
\end{tabularx}
\end{table}

\section{Summary and Open Questions}\label{sec:summary}
In summary, we highlight how the physics of pyrochlore iridates has been reinvigorated by moving beyond bulk crystals to epitaxially engineered thin films and heterostructures. The recent breakthrough in materials synthesis based on the solid phase epitaxy method allowed to overcome the significant synthesis challenges and enabled powerful new tuning knobs including dimensional confinement, strain, and interfacial coupling. The availability of high-quality films open the door for a host of emergent quantum phenomena, from magnetic Weyl semimetals to chiral spin liquid-like states. Building on these foundational achievements the field is poised to explore several exciting and yet challenging frontiers. In addition, the ability to manipulate these complex quantum materials at the atomic scale opens up a vast parameter space for discovering new physics. Here, we outline some of the key open questions and ideas for future research directions.

\subsection{New Materials to Fabricate}

The theoretical phase diagrams of pyrochlore iridates predict a multitude of exotic quantum states, many of which are delicately balanced on the edge of phase transitions \cite{goswami-2017-prb, witczak-krempa-2013-prb}. Accessing these novel phases, therefore, is not merely a matter of observation but requires the active efforts in fabrication of new material systems where the fundamental parameters such as lattice constant, bond angles, and dimensionality can be rigorously controlled. Future progress is thus deeply intertwined with advances in materials synthesis, which provides the foundation for all new discoveries \cite{chakhalian-2020-aplmater,kim-2022-aplmater, gutierrez2025epitaxial}.

\textit{Ultrathin films.}
The emergence of a chiral spin-liquid-like state in Y$_2$Ir$_2$O$_7$ films thinner than 30~nm vividly illustrates how dimensional confinement can reshape the magnetic landscape in pyrochlores \cite{liu-2024-natcommun}. In this quasi-2D limit, long-range magnetic order is destroyed and quantum fluctuations are markedly enhanced, allowing for the existence of novel entangled ground states. A key open question is whether this behavior represents a broader trend within the pyrochlore iridate family. Future work will involve pushing the boundaries towards atomic-layer-by-layer growth to fabricate pristine bilayer or tri-layer films \cite{guo_npjCM_2021}. Those ultra-thin material could stabilize novel chiral orders, a quantum Hall-like state or other entangled ground states theoretically predicted for the pyrochlore slab \cite{yang2014emergent, PhysRevLett.114.016806, kawamura-2002-prl}.

\textit{Heterostructures and interfaces.}
The interface between two different materials offers a fertile playground for engineering emergent phenomena through proximity effects. Such interfaces not only allow rigorous experimental tests of theoretical predictions but also open directions to discovering physics beyond the Landau paradigm \cite{chakhalian-2020-aplmater}. A central challenge in this field is the realization of theoretically proposed interfacial phases, for example, a two-dimensional gas of magnetic monopoles with net magnetic charge at a spin-ice/iridate heterojunction \cite{miao-2020-natcommun}. A critical prerequisite for achieving this is the fabrication of atomically sharp and structurally coherent interfaces. Recent breakthroughs in a hybrid solid-state epitaxy method have enabled the creation of pristine pyrochlore interfaces, such as between the spin ice Dy$_2$Ti$_2$O$_7$ and the Weyl semimetal Eu$_2$Ir$_2$O$_7$ \cite{kareev-2025-nanolett}. This new synthesis approach, which effectively uses the underlying pyrochlore crystal as an ideal template, minimizes defects and provides an unprecedented level of chemical  and structural control for designing complex heterostructures and superlattices.

\textit{Chemical doping.}
Controlled introduction of dopants into the pyrochlore matrix provides a direct way to apply chemical pressure and tune the electronic bandwidth and magnetic interactions. This approach has been explored in bulk crystals, for instance by substituting different rare-earth ions or applying high pressure to map the topological phase diagram and explore quantum critical points \cite{Ueda2016, porter-2019-prb, coak-2024-npjqm}. Applying the same principles to thin films remains unexplored and very promising area for exploration.

\subsection{New Physics to Explore and Experimental Probes}

While new materials provide the canvas, the ultimate challenge lies in identifying the novel physics painted upon it. Many of the most fascinating predicted phenomena in pyrochlores such as multipolar orders or the neutral excitations of a quantum spin liquid are considered `hidden' because they lack clear signatures in conventional thermodynamic or transport measurements. Moreover, majority of advanced scattering probes are dipole selective and as such many interesting new states remain `invisible'. Therefore, a crucial open frontier is the development and application of new experimental probes that can bring these subtle quantum states into view.

\textit{Topological surface states.}
While transport measurements offer indirect signatures of topological phases, direct visualization of the electronic structure is essential for unambiguous confirmation. In a WSM, this means observing the characteristic Fermi arcs, whereas for a proposed axion insulator it involves detecting gapped surface states with a distinct magnetic-field dependence.  Surface-sensitive probes like angle-resolved photoemission spectroscopy (ARPES) are the ideal tool for this task, and the theoretical groundwork for the expected surface signatures of pyrochlore axion insulators has been established, providing a clear roadmap for future experiments \cite{varnava-2018-prb}.

\textit{Domains, hidden phases, and multipolar orders.}
The direct, real-space visualization of magnetic structures is highly desired. The groundbreaking use of microwave impedance microscopy to visualize conductive domain walls in Nd$_2$Ir$_2$O$_7$ confirmed foundational theoretical predictions \cite{ma-2015-science, yamaji-2014-prx, ueda-2014-prb}. An open challenge is to image the AIAO spin texture within the domains. A particularly intriguing open question is the possibility of `hidden phases' that are invisible to bulk magnetic probes but can be revealed by local or transport measurements. For example, recent work on (Eu,Bi)$_2$Ir$_2$O$_7$ heterostructures grown on a spin ice has revealed a hidden magnetic transition deep within the AIAO ordered state, which manifests as an anomalous feature in the anisotropic magnetoresistance and is tunable by an external field \cite{wu-2025-sciadv}. Understanding the microscopic nature of such hidden transitions requires advanced local probes. Similarly, detecting hidden multipolar orders remains a major frontier. While the planar Hall effect has been used as a powerful indirect probe \cite{song-2022-natcommun}, the future may lie in using techniques with tensorial sensitivity, such as optical second harmonic generation, to directly detect the symmetry breaking associated with these complex magnetic structures \cite{harter-2017-science, suzuki-2019-prb, talanov-2023-actamater, fu-2015-prl, ladovrechis-2021-prb}.

\textit{Excitations.}
Understanding the elementary excitations is a key to characterizing any new phases of matter. This is in particular essential when the long-range order is suppressed due to enhanced quantum fluctuation at reduced dimensions, or significant frustration from multiple competing interactions. The macroscopic behaviors of the system may look trivial when probed by conventional probes, whereas the true nature of the system can only be revealed based on signatures from the excitation spectroscopy \cite{khomskii-2021-nc,smith-2022-prx, wen-2019-npjqm}.  For pyrochlore iridates, neutron scattering is limited by the strong neutron absorption of iridium. Resonant inelastic X-ray scattering (RIXS), therefore, has emerged as the leading technique for mapping momentum-resolved magnon dispersions across entire Brillouin zone, and remains critical for probing excitations in thin films and heterostructures. Complementary spectroscopies further enrich this picture. Surface enhanced Raman or Infrared spectroscopy, for instance, is a powerful table-top probe of local spin-phonon coupling and crystal field excitations \cite{son-2019-npjqm, PhysRevB.100.115157, lee-2013-prb, nguyen-2021-prl}. Time-domain terahertz (THz) spectroscopy can directly measure the low-energy optical conductivity and probe the collective response in search for axion electrodynamics, as demonstrated in the Luttinger semimetal Pr$_2$Ir$_2$O$_7$ \cite{cheng-2017-natcommun}. For charge neutral excitations in quantum disordered states like a spin liquid, the thermal Hall effect is a key experiment to detect unique signatures of spinons or topological magnons \cite{hwang-2020-prl,zhang-2020-prr}.

%Understanding the elementary excitations is key to characterizing any new phase of matter. This is in particular essential when the long-range order is suppressed due to enhanced quantum fluctuation at lower dimensions, or significant frustration from multiple competing interactions. The macroscopic behaviors of the system may look trivial probed by conventional probes, whereas the true nature of the system can only be revealed based on signatures from the excitation spectroscopy \cite{khomskii-2021-nc,smith-2022-prx, wen-2019-npjqm}. RIXS is the premier tool for mapping the momentum-resolved magnon dispersions of the AIAO state and remains crucial for studying excitations in thin films \cite{lee-2013-prb, donnerer-2019-jpcm}. This can be complemented by other spectroscopic techniques. Raman spectroscopy, for instance, is a powerful table-top probe of local spin-phonon coupling and crystal field excitations \cite{ishizuka-2019-prb, son-2019-npjqm, nguyen-2021-prl}. Time-domain terahertz (THz) spectroscopy can directly measure the low-energy optical conductivity and probe the collective electrodynamic response, as demonstrated in the Luttinger semimetal Pr$_2$Ir$_2$O$_7$ \cite{cheng-2017-natcommun}. For the neutral excitations expected in quantum disordered states like a spin liquid, the thermal Hall effect is a key proposed experiment to search for the unique signatures of spinons or topological magnons \cite{hwang-2020-prl}.

\subsection{Time-Dependent and Nonequilibrium Phenomena}

Beyond their static ground states, an emerging frontier lies in probing the non-equilibrium dynamics of pyrochlore iridates on their intrinsic ultrafast timescales. Driving these materials out of equilibrium with optical fields enables the creation of transient quantum states that are inaccessible in thermal equilibrium. This exciting venue of pump-probe spectroscopy and Floquet engineering represents an important shift from passively observing emergent phenomena to actively controlling them\cite{topp-2018-natcommun}.

Floquet engineering, in particular, can induce WSM phases in otherwise topologically trivial systems.  Under circularly polarized light time-reversal symmetry is broken and the electronic band structure can be drastically modified to generate Weyl nodes dynamically.
In relevance to pyrochlore iridates, this was suggested to be possible starting from the band insulators~\cite{zhang2016theory}, Dirac semimetals~\cite{hubener2017creating}, and Luttinger semimetals~\cite{ghorashi2018irridated}, where the resulting Weyl nodes can be further tuned by an applied magnetic field or by combined strain–field coupling~\cite{goswami2017competing,oh2018magnetic}.
%
%In relevance to pyrochlore iridates, this was suggested to be possible starting from the band insulators~\cite{zhang2016theory}, Dirac semimetals~\cite{hubener2017creating}, and Luttinger semimetals~\cite{ghorashi2018irridated}, where the resulting Weyl nodes can be further tuned via magnetic fields or a combination of strain and magnetic field~\cite{goswami2017competing,oh2018magnetic}.
To date, majority of Floquet-induced WSMs have been understood primarily within non-interacting or weakly correlated frameworks.  Realizing light-induced correlated WSMs in pyrochlore iridates, by contrast would provide access to a new regime in which topologically nontrivial phases emerge from the interplay of the Floquet drive and strong correlations.
%Importantly, Floquet-engineered Weyl semimetal phases in all materials accessed to date can be described using a non-interacting or weakly interacting picture.
%Realizing light-induced Weyl semimetals in pyrochlore iridates, by contrast, provides access to a new regime in which topologically nontrivial phases emerge from the interplay of the Floquet drive and strong correlations.

Furthermore, theoretically it has been proposed that a Mott insulator to WSM transition could be triggered by a single ultrafast laser pulse that transiently reduces the effective Hubbard interaction $U$. The emergence of such a transient Weyl phase and its chiral nodes could, in principle, be directly observed using time- and angle-resolved photoemission spectroscopy (tr-ARPES)\cite{yu2016determining,ma2017direct,yang-2017-sr}. Experimentally, however, photoemission studies on pyrochlore iridates remain very limited~\cite{wang2014experimental, kondo2015quadratic,nakayama2016slater}, primarily due to the challenge of preparing the large, clean surfaces required for such measurements~\cite{sobota2021angle}.

\funding{}
%The authors deeply acknowledge ...
X.L. and J.G. acknowledge the support by the National Natural Science Foundation of China (Grant No. 12434007, 12250710675) and the National Key R\&D Program of China (Grant No. 2022YFA1403400).
M.T. and J.C. acknowledge the support by the U.S. Department of Energy, Office of Science, Office of Basic Energy Sciences under award no. DE-SC0022160. M.T. also acknowledges the support by the Claud Lovelace Fellowship from the Department of Physics and Astronomy, Rutgers University.
A.-M.N. was supported by the U.S. Department of Energy (DOE) through the University of Minnesota (UMN) Center for Quantum Materials under Grant No. DE-SC0016371.

%\section*{References}
\bibliographystyle{unsrt}
\bibliography{refs_final}

%
% Each of the commands below will create an unnumbered section with the appropriate heading.
% Remove any sections that are not relevant for your article.
% All sections except suppdata will be removed if the [anonymous] option is used.
% See iopjournal-guidelines.pdf for more information.
%

%\ack{Sample text inserted for demonstration.}

%\funding{Sample text inserted for demonstration.}
% This section is a list of funder names and grant numbers

%\roles{Sample text inserted for demonstration.}
% List author names and the contributions made to the article, using terms from the NISO Contributor Roles Taxonomy (CRediT) https://credit.niso.org

%\data{Sample text inserted for demonstration.}
% For more information on IOP Publishing's research data policy see: https://publishingsupport.iopscience.iop.org/questions/research-data/

%\suppdata{Sample text inserted for demonstration.}

\end{document}